\documentstyle[preprint,aps]{revtex}
\tighten
\begin{document}
\draft
\preprint{FSU-SCRI-94-11}
\title{
Metastable lifetimes in a kinetic Ising model:
Dependence on field and system size
}
\author{
Per~Arne~Rikvold,$^{1,2,3,}$* H.~Tomita,$^1$ S.~Miyashita,$^2$
and Scott~W.\ Sides$^3$
}
\address{
$^1$Department of Fundamental Sciences, Faculty of Integrated Human Studies,
Kyoto University, Kyoto 606, Japan \\
$^2$Graduate School of Human and Environmental Studies,
Kyoto University, Kyoto 606, Japan \\
$^3$Supercomputer Computations Research Institute,
Center for Materials Research and Technology, and Department of Physics, \\
Florida State University, Tallahassee, Florida, 32306-4052\dag
}
\date{\today}
\maketitle
\begin{abstract}
The lifetimes of metastable states in kinetic Ising ferromagnets
are studied by droplet theory and Monte Carlo simulation,
in order to determine their dependences on applied field and system size.
For a wide range of fields, the dominant field dependence
is universal for local dynamics and has the form of
an exponential in the inverse field, modified by
universal and nonuniversal multiplicative power-law prefactors. Quantitative
droplet-theory predictions for these dependences are numerically verified,
and small deviations from the predictions
are shown to depend nonuniversally on the details of the dynamics.
We identify four distinct field intervals in which the field dependence
and statistical properties of the lifetimes are markedly different.
The field marking the crossover between the weak-field regime,
in which the decay is dominated by a single droplet, and the
intermediate-field regime, in which it is dominated by a finite density of
droplets, vanishes logarithmically with system size. As a consequence the
slow decay characteristic of the former regime may be
observable in systems that are macroscopic as far as their equilibrium
properties are concerned.
\end{abstract}
\narrowtext

\section{Introduction}
\label{sec1}

Although it is observed in nature in contexts as
dissimilar as supercooled water or water vapor
\cite{MCDO62,ABRA74,water} and the electroweak \cite{eweak} and
QCD confinement/deconfinement \cite{QCD} phase transitions,
metastability is very difficult to characterize in a
microscopically precise fashion \cite{SCHU90}.
In certain systems with weak long-range
interactions infinitely long-lived metastable states can
exist in the thermodynamic limit \cite{PENR71},
and metastability in such systems has been studied with several techniques
\cite{GOFI94}, including
field theoretical, Monte Carlo (MC), and transfer-matrix methods.
However, in systems with
short-range interactions metastable states eventually decay,
even though their lifetimes may be many orders of magnitude larger than
other characteristic timescales of the system, and may even become comparable
to the age of the universe \cite{MCDO62}.
In order to gain a deeper understanding of metastability in short-range
systems, it is necessary to consider in detail the
lifetimes of metastable states and how they are determined by the physical
mechanisms involved in the decay \cite{FISH90}.

As a prototype for the metastable
dynamics of short-range systems, the decay of the magnetization in
impurity-free kinetic Ising ferromagnets in unfavorable applied fields
has been studied by MC methods in both two
\cite{STOL72,BIND73A,BIND73B,BIND74,STOL77,MCCR78,PAUL88,RAY90A,BIND76}
and three \cite{STAU82,RAY90B,STAU92} dimensions. The results of
several of these studies were analysed in terms of droplet theory
\cite{FISH67,LANG67,LANG68,LANG69,GNW80,GUNT83,PENR94}, establishing
general agreement between theory and simulations.
In the present work we further investigate the extent of that
agreement by analytical droplet-theory calculations and MC simulations.
We emphasize the dependences on applied field and
system size of the metastable lifetimes
and their relations to particular decay mechanisms.
Our analysis extends a recent study by two of us
\cite{TOMI92A,TOMI92B}, and Ref.~\cite{TOMI92A} is
henceforth referred to as (I).

The main results in (I) can be summarized as follows.
In agreement with previous \cite{STOL77,RAY90B}
and more recent \cite{STAU92} simulations it was found
that the statistical properties and system-size
dependence of the metastable lifetimes in MC simulations
were markedly different in two
separate field regions. The point separating these regions was called
``the dynamic spinodal point'' (DSP), and the corresponding field,
which depends on temperature and system size, was denoted $H_{\rm DSP}$.
For $|H| \! > \! H_{\rm DSP}$, the mean
lifetime was observed to be independent of system size
and much greater than its standard deviation \cite{STOL77,RAY90B,STAU92}.
This field region was therefore called ``the deterministic region''.
For weaker fields, the mean lifetime was inversely
proportional to the system volume
\cite{STOL77,MCCR78,RAY90B,STAU92,LANG67,LANG68,LANG69},
and the standard deviation was
approximately equal to the mean \cite{STOL77,RAY90B,STAU92}.
For these reasons this regime was termed ``the stochastic region''.
The position of the DSP, which separates these regions, depends on the
temperature and system size. The main goal of the present work
is to understand and quantify these dependences.

In this paper
we present a quantitative analysis of metastable lifetimes in terms of
a droplet model of homogeneous nucleation in $d$-dimensional systems
\cite{FISH67,LANG67,LANG68,LANG69,GNW80,GUNT83,PENR94},
obtaining analytic results for the logarithmic derivative
of the lifetime with respect to $1/|H|^{d-1}$, as well as the explicit
size and temperature dependence of $H_{\rm DSP}$.
Both the deterministic and the stochastic region are found to be divided into
two subregions, and expressions for the corresponding crossover
fields are derived. These analytic results are shown to be
in excellent agreement with MC simulations
for $d$=2 at a temperature of 0.8$T_{\rm c}$.

In order of increasingly strong unfavorable field the distinct regions
that we identify are:
the ``coexistence region'', characterized
by subcritical fluctuations on the scale of the system volume; the
``single-droplet region'', characterized by decay via a single critical
droplet; the ``multi-droplet region'', characterized by decay via a finite
density of droplets; and the ``strong-field
region'', in which the droplet picture is inappropriate.
The two former regions comprise the stochastic region, and the two latter ones
the deterministic region. The DSP thus marks the
crossover between the single-droplet and multi-droplet regions.
The crossover between the coexistence region and the single-droplet region
was called ``the thermodynamic spinodal point'' (THSP) in (I),
and the crossover between the
strong-field and multi-droplet regions  was called
``the mean-field spinodal point'' (MFSP).
These different regions and the crossover points that separate them
are summarized in Fig.~\ref{fig0}, which shows a schematic sketch of the
field dependence of the average metastable lifetime.

Whereas the dominant field dependence of the lifetime
in the single-droplet and multi-droplet regions
is exponential in the inverse field,
our MC data are sufficiently precise to also allow evaluation of
multiplicative power-law prefactors.
These are predicted by droplet theory
\cite{LANG67,LANG68,LANG69,GNW80,GUNT83}
but are difficult to detect numerically and, to our knowledge, have not
previously been measured in MC studies of lifetimes or nucleation rates.

The rest of the paper is organized as follows.
In Sec.~\ref{sec2} the kinetic Ising model is defined, and the
numerical methods used in this work
are discussed. In Sec.~\ref{sec3} the droplet-theoretical
predictions are developed, for both infinitely large and finite systems.
In Sec.~\ref{sec4} the numerical results are presented and compared with
the theory, and Sec.~\ref{sec5} contains a brief discussion.

\section{Model and numerical methods}
\label{sec2}

The model is defined by the reduced Hamiltonian,
\begin{equation}
\label{eq1}
{\cal H}/k_{\rm B}T = - K \sum_{\langle i,j \rangle} s_i s_j - H \sum_i s_i
\;,
\end{equation}
where $s_i \! = \! \pm1$ is the spin at site $i$,
$T$ is the temperature, $k_{\rm B}$ is Boltzmann's constant,
$K \! = \! J/k_{\rm B}T$ ($>$0) and $H \! = \! h/k_{\rm B}T$
are the reduced coupling constant and field, respectively. The sums
$\sum_{\langle i,j \rangle}$ and $\sum_i$ run over all nearest-neighbor pairs
and over all $N$ sites on a $d$-dimensional hypercubic lattice, respectively.

The dynamics is given by the Metropolis single-spin-flip MC
algorithm. The transition probability for a flip of
the spin at site $\alpha$ from $s_\alpha$ to $- \! s_\alpha$ is
thus defined as \cite{MKB73}
\begin{mathletters}
\label{eq1a}
\begin{equation}
\label{eq1aa}
W( s_\alpha \rightarrow - \! s_\alpha ) =
\min [ 1 ,  \exp (- \! \Delta E_\alpha) ] \;,
\end{equation}
where
\begin{equation}
\label{eq1ab}
\Delta E_\alpha = 2 s_\alpha \left[ K \sum_{\rm NN} s_\beta + H \right]
\;
\end{equation}
\end{mathletters}
is the energy change during the flip, divided by $k_{\rm B} T$.

We study the relaxation of the magnetization,
\begin{equation}
\label{eq1b}
m(t) = N^{-1} \sum_{i=1}^N s_i(t) \;,
\end{equation}
starting from an initial state magnetized opposite to the applied field.
This approach was also used, {\it e.g.}\ in Ref.~\cite{STAU92}.
The magnetization $m(t)$
is directly obtained as an average
over the droplet size distribution \cite{BIND76,STAU92,FISH67},
and it is closely related to the
nonequilibrium relaxation functions introduced by Binder \cite{BIND73A}.
The volume fraction of stable phase at time $t$ is $\phi_{\rm s}(t)
= \left( m_{\rm ms} \! - \! m(t) \right) /
\left( m_{\rm ms} \! - \! m_{\rm s} \right)$,
where $m_{\rm s}$ and $m_{\rm ms}$ are the bulk equilibrium and metastable
magnetizations, respectively.

The simulations were performed on
square $L \! \times \! L$ systems, with $L$=64, 128,
256, 400, and 720 and periodic boundary conditions. The systems
were initialized with $m(0)$=+1 and allowed to develop
in a constant field $H$$<$0 at $T \! = \! 0.8T_{\rm c}$
($K \! = \! 0.550 \, 858...$).
This temperature is high enough to obtain
reasonable MC acceptance rates, but it is sufficiently low
to avoid complications such as critical slowing-down and possible
finite-size scaling effects due to small droplets \cite{MON87}.
For later reference we note that
the equilibrium magnetization at $0.8T_{\rm c}$
is $m_{\rm s} \! = \! 0.954 \, 410...$\, \cite{YANG52}, and the
equilibrium surface tension along a
primitive lattice vector is $\sigma_0 \! = \! 0.745 \, 915...$\, \cite{ZIA82}.

Average first-passage times
to $m$=0.9,~0.8, \ldots,~0.0 were recorded, together with their empirical
standard deviations, and the mean lifetime of the metastable state
was estimated as the average first-passage
time to $m$=0.7. We used this
particular cutoff magnetization for the following reasons.
In order for the droplet-theoretical considerations developed
in Sec.~\ref{sec3} to remain valid during the whole time evolution until
the cutoff, the cutoff value of $\phi_{\rm s}$ should be
below the percolation limit. However,
the cutoff magnetization should not be so close to $m_{\rm ms}$ that
subcritical fluctuations in the metastable phase
(``recrossing events'' \cite{PAUL88}) might be mistaken for
decay events, thus leading to significant
underestimation of the mean lifetime. The cutoff magnetization $m$=0.7,
which, assuming $m_{\rm ms} \! = \! - \! m_{\rm s}$,
corresponds to $\phi_{\rm s} \! \approx \! 0.13$, was chosen as
a reasonable compromise.
However, except for very strong and very weak fields,
the dependence of the observed lifetimes
on the cutoff is weak, and results almost
identical to those reported here were obtained with cutoff at $m$=0.0,
or $\phi_{\rm s} \! \approx \! 0.5$.

A more accurate, but also more computationally intensive, method to estimate
the lifetimes would be to
use the recrossing events by choosing as a field-dependent
cutoff a value of $m$ which lies on the side of the maximum of
the recrossing-event distribution \cite{PAUL88}
opposite from that of the initial magnetization, and
for which the recrossing-event probability is below
a given threshold. We do not
implement this method here, but plan to do so in future studies.

For fields at which the lifetime is large, the magnetization $m(t)$
was measured after each MC step per site (MCS), whereas for shorter lifetimes
it was measured after each successfull spin flip. The latter method was used
whenever necessary in order to
ensure that the uncertainty due to the discreteness of
the time variable was at least one order of magnitude smaller than the
standard deviations in the average first-passage times.

The numerical data were obtained
with a special-purpose {\it m}-TIS2 computer \cite{ITO87,ITO92}
at Kyoto University and on heterogeneous clusters of
IBM RS/6000, DEC-station, and DEC~3000 Model~400 Alpha
workstations at Florida State University.

The {\it m}-TIS2 architecture
only allows the MC updates to be performed sequentially,
and the available memory limited the maximum system size to $L$=720.
To assess the effects of the former restriction, simulations with
updates at randomly chosen sites
were performed on the workstation clusters for $L$=128 and~720.
The total computer time spent
was approximately 400 {\it m}-TIS2 hours and
1600 hours of workstation time for the sequential-update study,
and an additional 8000 workstation hours for the random-update investigation.

\section{Droplet theory}
\label{sec3}

In this section we derive the droplet-theoretical relations necessary to
interpret the simulated lifetime data.

The system is characterized by six length scales: the lattice constant,
which we take as unity; the single-phase correlation lengths
in the stable and metastable phases, $\xi_{\rm s}$ and $\xi_{\rm ms}$,
respectively;
the critical droplet radius $R_{\rm c}$; the size $R_0$
to which one droplet can grow before it is likely to meet another,
which we shall refer to as the mean droplet separation;
and the system size $L$.
Here we only consider cases in which $\xi_{\rm s}$ is smaller than
the other length scales, {\it i.e.}\ well below the critical temperature.
Thus we are left to consider the interplay between four lengths:
$L$, $R_0$, $R_{\rm c}$, and $\xi_{\rm ms}$.

\subsection{Infinite systems}
\label{sec3a}

First we obtain $R_0$ and $R_{\rm c}$ for $d$-dimensional systems in the limit
$L \! \rightarrow \! \infty$.
Although the droplets are almost circular at our simulation temperature,
we give a general argument
that remains valid for low temperatures, at which
nonspherical droplets appear because of the anisotropy of the surface
tension \cite{ZIA82,ROTT81,AVRO82,HARR84,CCAG93,CCAG94}.
The free energy of a (regular, but not necessarily spherical)
$d$-dimensional droplet of radius $R$
(defined as half the extent of the droplet along a primitive lattice vector)
and volume $V(R) \! = \! \Omega_d R^d$ is
\begin{eqnarray}
\label{eq5}
F(R) &=& V^{(d-1)/d}\widehat{\Sigma} - V \Delta \nonumber \\
     &=& \Omega_d^{(d-1)/d}  R^{d-1} \widehat{\Sigma} - \Omega_d R^d \Delta \;,
\end{eqnarray}
where $\Delta$ is the difference in bulk free-energy density between the
metastable and stable states.
The quantity ${\widehat{\Sigma}}$ is a temperature- and, in principle,
field-dependent proportionality factor which relates the surface
contribution to $F(R)$ with the droplet volume \cite{ZIA82,ZIA82X}.
Applying standard droplet-theory arguments \cite{GUNT83} to
$F(R)$ one finds the critical radius, the free-energy cost of a
critical droplet, and the nucleation rate per unit time and volume.
The critical radius is
\begin{equation}
\label{eq2a}
R_{\rm c}(T,H) = \frac{(d \! - \! 1)}{\Delta}
                 \frac{\widehat{\Sigma}}{d \Omega_d^{1/d}}
               = \frac{(d \! - \! 1)\sigma_0}{\Delta}
        \approx \frac{(d \! - \! 1)\sigma_0}{2 m_{\rm s} k_{\rm B}T|H|} \;,
\end{equation}
where $\sigma_0 \! = \! \widehat{\Sigma} / ( d \Omega_d^{1/d} ) $ is the
surface tension along a primitive lattice vector \cite{ZIA82}.
The approximation $\Delta \! \approx \! 2 m_{\rm s} k_{\rm B}T|H|$,
where $m_{\rm s}(T)$ is the spontaneous equilibrium magnetization, is
expected to be valid not only at low temperatures, but
even near the critical point \cite{HARR84}.
The free-energy cost of a critical droplet is
\begin{eqnarray}
\label{eq3}
F_{\rm c}(T,H)
&=& \left( \frac{d \! - \! 1}{\Delta} \right)^{d-1}
  \left( \frac{\widehat{\Sigma}}{d} \right)^d \nonumber \\
&\approx& \left( \frac{d \! - \! 1}{2m_{\rm s} k_{\rm B}T|H|} \right)^{d-1}
  \left( \frac{\widehat{\Sigma}}{d} \right)^d \;,
\end{eqnarray}
and the nucleation rate per unit time and volume, $I(T,H)$,
is determined by $F_{\rm c}$ through \cite{LANG67,LANG68,LANG69,GNW80}
\begin{mathletters}
\label{eq4}
\begin{equation}
\label{eq4a}
I(T,H) = A(T) |H|^{b+c} e^{ - \frac{F_{\rm c}(T,H)}{k_{\rm B} T} }
\approx A(T) |H|^{b+c} e^{- \frac{ \Xi }{ |H|^{d-1}} }
\end{equation}
with
\begin{equation}
\label{eq4b}
\Xi =
\left( \frac{d \! - \! 1}{2m_{\rm s}}\right)^{d-1}
\left( \frac{\widehat{\Sigma}}{d k_{\rm B} T}\right)^d
\;,
\end{equation}
\end{mathletters}
where the approximation introduced in the second part of Eq.~(\ref{eq4a}) is
the same as in Eqs.~(\ref{eq2a}) and~(\ref{eq3}).
The function $A(T)$ is expected to be nonuniversal, $b$ is a universal
exponent related to
Goldstone-mode ``wobbles'' on the droplet surfaces \cite{LANG67,GNW80},
and $c$ gives the $H$ dependence of the ``kinetic prefactor''
\cite{LANG68,LANG69},
which is the only part of $I(T,H)$ that may depend explicitly on the
specific dynamics. A field-theoretical calculation gives \cite{GNW80}
\begin{equation}
\label{eq14}
b = \left\{ \begin{array}{ll} (3 \! - \! d)d/2 &
                                           \mbox{~for $1 \! < \! d \! < \! 5$,
                                                    $d \! \neq \! 3$} \\
                              -7/3     & \mbox{~for $d$=3}
            \end{array}\right.
\;.
\end{equation}
Strictly speaking, a multiplicative correction term of form
$[1 \! + \! O(H^2)]$ should be included in the exponent in
Eq.~(\ref{eq4a}) \cite{GNW80,HARR84,CCAG93,CCAG94,JACU83,PERI84},
but its effect on the
metastable lifetimes is small for the relatively weak fields on which we focus
our attention in this work, and it has therefore been suppressed for
simplicity.

For $d$=2 there is substantial numerical evidence
that $b$=1, as predicted by Eq.~(\ref{eq14}). This is obtained from
calculations that do not involve the dynamics,
such as analyses of series expansions
\cite{HARR84,LW80,WALL82} and transfer-matrix calculations
\cite{CCAG93,CCAG94}.
These studies, as well as MC work \cite{JACU83,PERI84}, also indicate
that the free-energy cost of the critical droplet
is given by Eq.~(\ref{eq3}) with the zero-field equilibrium value for
$\widehat{\Sigma}$. We therefore adopt the
notations $\widehat{\Sigma}(T)$ and $\Xi(T)$ to emphasize the lack of
field dependence in these quantities. They can be obtained
with arbitrary numerical precision by combining a Wulff construction
with the exact, anisotropic zero-field surface tension \cite{ZIA82}.
This general result, that the surface free energy of compact critical
droplets is determined by the zero-field equilibrium surface tension, is also
supported by MC studies of nucleation rates in three dimensions
\cite{STAU82,RAY90B,STAU92}.
For dynamics that can be described by a Fokker-Planck equation it is expected
that the kinetic prefactor is proportional to $R_{\rm c}^{-2}$
\cite{LANG68,LANG69,GNW80}, which by Eq.~(\ref{eq2a})
would yield $c$=2. However, we are not aware that
independent, numerical verifications of this result have been performed
previously. As discussed in Sec.~\ref{sec4} we find that the measured
value of $c$ depends on the MC update scheme.

Assuming that the growing droplets are not substantially deformed
\cite{BREN92}, the radial growth velocity is obtained
in an Allen-Cahn approximation \cite{GUNT83,ALLE79,SEKI86} as
\begin{equation}
\label{eq6}
v_\bot = (d \! - \! 1) \Gamma \left( R_{\rm c}^{-1} \! - \! R^{-1} \right)
\stackrel{\scriptscriptstyle R \rightarrow \infty}{\longrightarrow}
(d \! - \! 1) \Gamma R_{\rm c}^{-1} \equiv v_0 \;,
\end{equation}
where $\Gamma$ depends on the details of the kinetics.
The mean droplet separation $R_0$ and its associated time scale $t_0$ are
obtained from $v_\bot$ and the nucleation rate $I$ by requiring that
$R_0 \! = \! v_\bot t_0$ and $R_0^d t_0 I \! = \! 1$ \cite{SEKI86}.
If $R_0 \! \gg \! R_{\rm c}$,
so that $v_\bot \! \approx \! v_0$, then by using
$v_0 \! \propto \! |H|$ from Eqs.~(\ref{eq2a}) and~(\ref{eq6}), one obtains
\begin{mathletters}
\label{eq7}
\begin{equation}
\label{eq7a}
t_0(T,H) = (v_0^d I)^{- \frac{1}{d+1}}
         = B(T) |H|^{- \frac{b+c+d}{d+1}} e^{ \frac{1}{d+1}
                       \frac{ \Xi (T) }{ |H|^{d-1}}} \;,
\end{equation}
and
\begin{equation}
\label{eq7b}
R_0(T,H) = v_0 t_0
         = C(T) |H|^{- \frac{b+c-1}{d+1}} e^{ \frac{1}{d+1}
                       \frac{ \Xi (T) }{ |H|^{d-1}}} \;.
\end{equation}
\end{mathletters}
The coefficients $B$ and $C$ have the same nonuniversal status as $A$ in
Eq.~(\ref{eq4a}).
Although
$R_{\rm c} \! \sim \! |H|^{-1} \! \rightarrow \! \infty$ as
$|H| \! \rightarrow \! 0$,
$R_{\rm c}/R_0 \! \rightarrow \! 0$ in the same limit.

If one assumes that the positions of the (possibly overlapping)
growing droplets are uncorrelated, the volume fraction
of stable phase at time $t$ becomes
\begin{equation}
\label{eq9x}
\phi_{\rm s}(t) = 1 - \exp \left[ { - \frac{\Omega_d}{d \! + \! 1}
\left( \frac{t}{t_0} \right)^{d+1} } \right] \;,
\end{equation}
which is known as Avrami's law  \cite{SEKI86,KOLM37,JOHN39,AVRA39}.
{}From this it is seen that the average time
taken to reach a specific $\phi_{\rm s}$ is
\begin{equation}
\label{eq9}
\langle t(\phi_{\rm s}) \rangle
= t_0(T,H)
  \left[ - \frac{d \! + \! 1}{\Omega_d}
            \ln (1 \! - \! \phi_{\rm s}) \right]^{1/(d+1)} \;.
\end{equation}
The ``ideal-gas'' approximation leading to Avrami's law
is expected to hold when the total volume
fraction of droplets is sufficiently small that
droplet-droplet correlations can be
ignored. The intermediate-field region, in which
Eq.~(\ref{eq9}) holds, we call {\it the multi-droplet region}. It corresponds
to the picture of decay through continuously nucleating and growing
droplets, first introduced by Kolmogorov, Johnson and Mehl, and Avrami
over fifty years ago \cite{SEKI86,KOLM37,JOHN39,AVRA39}.

For stronger fields, a picture based on localized droplets no longer
adequately describes the system, which in that case
decays via long-wavelength,
unstable modes reminiscent of spinodal decomposition \cite{GUNT83}.
The crossover field separating the multi-droplet region from this
{\it strong-field region} was termed ``the mean-field spinodal'' in (I).
It is located at a field $H_{\rm MFSP} \! < \! 4K$
\cite{TOMI92B,NEVE91,SCHO92}. We expect that $R_{\rm c} \! < \! R_0$ for
all $H \! < \! H_{\rm MFSP}$.
A recent argument based on a transfer-matrix calculation indicates that
the single-phase correlation length in the metastable phase is
given by $\xi_{\rm ms} \approx T/(4J \! - \! 2h)$ at low $T$ \cite{CCAG94}.
A rough estimate for the field at which droplet theory breaks down can
be obtained by requiring that $2R_{\rm c} \! = \! \xi_{\rm ms}$.
This yields $H_{\rm MFSP} \! < \! 2K$ and specifically
$H_{\rm MFSP}(0.8T_{\rm c}) \! \approx \! 0.33$ with the critical-droplet
diameter $2R_{\rm c} \! \approx \! 1.3$, independent of system size.
The dynamics in the strong-field region beyond $H_{\rm MFSP}$
will not be discussed in detail here.

\subsection{Finite-size effects}
\label{sec3b}

Next we study the effects produced by a finite system size $L$.
For simplicity we consider hypercubic systems of volume $L^d$ with periodic
boundary conditions. Effects of heterogeneous nucleation at
free boundaries are discussed in Ref.~\cite{TOMI93}.
In the large-$L$ limit,
\begin{equation}
\label{eq10}
L \gg R_0 \gg R_{\rm c} \;,
\end{equation}
we use an approximate argument by which the system is partitioned into
$(L/R_0)^d \! \gg \! 1$ cells of volume $R_0^d$. Each cell decays from the
metastable state to the equilibrium state in an independent Poisson process of
rate $R_0^d I \! = \! t_0^{-1}$. The volume fraction is then self-averaging,
with $\langle t(\phi_{\rm s}) \rangle$
approximately equal to the infinite-system result of Eq.~(\ref{eq9}), and
the relative standard deviation is
\begin{equation}
\label{eq10b}
r = \frac{\sqrt{\langle t(\phi_{\rm s})^2 \rangle -
          \langle t(\phi_{\rm s}) \rangle^2 }}
         {\langle t(\phi_{\rm s}) \rangle}
  \approx (R_0/L)^{\frac{d}{2}} \rho    \;,
\end{equation}
where $\rho \! \approx \! 1$ is the relative standard deviation of a single
Poisson process.
The regime characterized by $r \! \ll \! 1$ was
termed ``the deterministic region'' in (I). It is subdivided
into the multi-droplet and strong-field subregions discussed above.

For smaller $L$, so that
\begin{equation}
\label{eq11}
R_0 \gg L \gg R_{\rm c} \;,
\end{equation}
the random nucleation of a single critical droplet in a Poisson process
of rate $L^d I$ is the rate-determining step. This is
followed by rapid growth, until this droplet occupies the entire system after
an additional time much shorter than the average waiting time before a
second droplet nucleates.
Therefore, in this case the characteristic lifetime is
\begin{eqnarray}
\label{eq12}
\langle t(\phi_{\rm s}) \rangle
&\approx& \left( L^d I(T,H) \right)^{-1} \nonumber \\
&\approx& L^{-d} [A(T)]^{-1} |H|^{-(b+c)} e^{\frac{ \Xi (T) }{ |H|^{d-1}} }
\end{eqnarray}
with $r \! \approx \! 1$ and
only a weak dependence on the threshold $\phi_{\rm s}$.
This single-droplet region is part of ``the stochastic region'' observed in
(I).

The crossover between the deterministic and stochastic regimes is
determined by the condition $L \! \propto \! R_0$. We identify the
crossover field with the ``dynamic spinodal field'' introduced in (I), and
in the limit $H \! \rightarrow \! 0$ we
explicitly obtain from Eq.~(\ref{eq7b})
\begin{equation}
\label{eq15}
H_{\rm DSP} =
              \left( \frac{1}{d \! + \! 1}
                     \frac{\Xi(T)}{\ln L}\right)^{\frac{1}{d-1}}
\left( 1 + O\left( \frac{\ln ( \ln L) }{\ln L } \right)
         + O\left( \frac{1}{\ln L } \right) \right) \; .
\end{equation}
We emphasize that, although $H_{\rm DSP}$ vanishes as
$L \! \rightarrow \! \infty$, the approach to zero
is exceedingly slow, especially for $d$=3 and larger. Therefore,
$H_{\rm DSP}$ may well be measurably different from zero for
systems that are definitely macroscopic as far as their equilibrium
properties are concerned.
(As an illustration, increasing $L$ from 100 to 10$^{10}$ for $d$=3 decreases
the leading term in
$H_{\rm DSP}$ only to approximately one-half of its original value!)
Recent exact results \cite{NEVE91,SCHO92} indicate that
if the limit $T \! \rightarrow \! 0$ is taken {\it before}
$L \! \rightarrow \! \infty$, then $H_{\rm DSP}$ should remain nonzero.
An estimate for the mean lifetime at $H_{\rm DSP}$ in the infinite-$L$ limit
is obtained by setting
\begin{eqnarray}
\label{eq15a}
\langle t \; {\rm at} \; H_{\rm DSP} \rangle & \propto & t_0(T,H_{\rm DSP})
= \frac{R_0(T,H_{\rm DSP})}{v_0(T,H_{\rm DSP})} \nonumber \\
 & \propto & L / H_{\rm DSP} \sim L (\ln L)^{1/(d-1)} \;,
\end{eqnarray}
where the nonuniversal, temperature dependent proportionality factors
have been dropped.

Finally, we consider the limit
\begin{equation}
\label{eq16}
R_0 \gg R_{\rm c} \gg L \;.
\end{equation}
In this case the volume term can be neglected in Eq.~(\ref{eq5}), and the
free-energy cost of a droplet occupying a volume fraction
$\phi_{\rm s} \! = \! V(R)/L^d$ is
$F(\phi_{\rm s}) \! = \! L^{d-1} \phi_{\rm s}^{(d-1)/d} \widehat{\Sigma}(T)$,
so that the first-passage time to a given $\phi_{\rm s}$
is independent of $H$ and diverges exponentially with
$L^{d-1}$.
Since the dynamics in this region of extremely weak fields
or extremely small systems is similar to that
on the coexistence line, $H$=0, we call it {\it the coexistence region}.
The crossover field between the coexistence and single-droplet regions,
called ``the thermodynamic spinodal point''
(THSP) in (I), is determined for a given $\phi_{\rm s}$ by
$\Omega_d (R_{\rm c}/L)^d \! = \! \phi_{\rm s}$, which yields
\begin{equation}
\label{eq17}
H_{\rm THSP} = \frac{1}{L \phi_{\rm s}^{1/d}}
\frac{(d \! - \! 1) \widehat{\Sigma}(T)}{2 d k_{\rm B} T m_{\rm s}} \;.
\end{equation}

\section{Numerical results}
\label{sec4}

The mean metastable lifetimes are illustrated in
Fig.~\ref{fig1}, which shows the average first-passage time to $m$=0.7
($\phi_{\rm s} \! \approx \! 0.13$)
for two-dimensional, square Ising systems with
$L$=64,\ldots,~720 at 0.8$T_{\rm c}$.
The simulations were performed using the Metropolis single-spin-flip MC
algorithm with sequential updates, as discussed in Sec.~\ref{sec2}.
The lifetimes, in units of MCS,
are plotted on a logarithmic scale {\it vs.}\ $1/|H|$.
Each data point is averaged over $n$=100 independent realizations,
and the errors are calculated from the empirical standard deviation
$\sigma$ in $t(m$=0.7) as $\sigma / \sqrt{n}$.
For given $|H|$ and sufficiently small $L$ the lifetimes decrease
with increasing $L$ as
$\langle t(\phi_{\rm s}) \rangle \! \propto \! L^{-2}$, in
agreement with Eq.~(\ref{eq12}) for the single-droplet region and as shown
in Fig.~3 of (I). For larger $L$ they converge to an
$L$-independent value, as expected from Eq.~(\ref{eq9}) for the
multi-droplet region.
This $L^{-d}$ behavior was previously remarked in
Refs.~\cite{STOL77,MCCR78,RAY90B,STAU92}, as was the presence of the
$L$-independent regime in Refs.~\cite{STOL77,RAY90B,STAU92}.
The size-independent data in the intermediate- and strong-field
portions of Fig.~\ref{fig1} belong to the deterministic region,
whereas the size-dependent data in the weak-field,
long-time portion lie in the stochastic region.
The droplet-theory results of Eqs.~(\ref{eq7a}), (\ref{eq9}), and~(\ref{eq12})
show that in {\it both} the multi-droplet and
the single-droplet regions the field dependence of the lifetime
is determined by the nucleation rate $I(T,H)$, so that
the asymptotic slopes of $\ln \langle t(\phi_{\rm s}) \rangle$
{\it vs.}\ $1/|H|^{d-1}$ in the two regimes
are related by a factor $(d \! + \! 1)$.
Based on the four points for $L$=64 in the single-droplet region of
Fig.~\ref{fig1} ($1/|H| \! \ge \! 20$) the effective slope
was estimated by a least-squares fit to be 0.36(1).
This slope is indicated by the dashed
line above the data in Fig.~\ref{fig1}, whereas the dashed line below the
data has 1/3 this slope. The agreement with droplet theory seems
excellent. However, the measured effective slope is almost 30\% larger
than the theoretically predicted asymptotic slope,
$\Xi(0.8T_{\rm c}) \! = \! 0.278 \, 840...$\,,
which is obtained by using the numerically exact \cite{ZIA82,ZIA82X} value
$\widehat{\Sigma}(0.8T_{\rm c})/J \! = \! 2.648 \, 81...$ in Eq.~(\ref{eq4b}).

The explanation for this apparent discrepancy lies in the power-law
prefactors in Eqs.~(\ref{eq7a}) and~(\ref{eq12}), which yield
the effective slopes
\begin{equation}
\label{eq13}
\Lambda_{\rm eff}
\equiv \frac{{\rm d} \, \ln \langle t(\phi_{\rm s}) \rangle}
            {{\rm d} \, (1/|H|^{d-1})}
= \lambda |H|^{d-1} + \Lambda \;,
\end{equation}
with $\lambda \! = \! (b \! + \! c)/(d \! - \! 1)$ and
$\Lambda \! = \! \Xi(T)$
in the single-droplet region described by Eq.~(\ref{eq12}), and
$\lambda \! = \! (b \! + \! c \! + \! d)/(d^2 \! - \! 1)$
and $\Lambda \! = \! \Xi(T)/(d \! + \! 1)$
in the multi-droplet region described by Eqs.~(\ref{eq7a}) and~(\ref{eq9}).

In Fig.~\ref{fig2} we show
two-point finite-difference estimates for $\Lambda_{\rm eff}$, based on the
average lifetimes in Fig.~\ref{fig1},
for pairs of fields, $|H_1| \! > \! |H_2|$. These
estimates are plotted {\it vs.}\ the
field value that occurs in the finite-difference version of Eq.~(\ref{eq13}),
$|H| \! = \! | H_1 H_2 | \ln | H_1/H_2 | / |H_1 \! - \! H_2|$.
The error bars are calculated as
$$
\sigma_\Lambda =
\sqrt{\left[ \left(\frac{\sigma_1}{\langle t_1 \rangle} \right)^2 +
             \left(\frac{\sigma_2}{\langle t_2 \rangle} \right)^2 \right]
             \frac{1}{n}} \left| \frac{H_1 H_2}{H_1 \! - \! H_2} \right| \;,
$$
where $\sigma_i$ and $\langle t_i \rangle$ are the empirical standard
deviation and mean for the lifetime at $|H_i|$, with $i$=1 or~2.
For clarity only two system sizes, $L$=128 and~720, were included in the
figure. The other sizes give results similar to those shown.
The solid straight lines correspond to Eq.~(\ref{eq13})
with $b \! + \! c \! = \! 2$ and the exact
$\Xi(0.8T_{\rm c}) \! = \! 0.278 \, 840...$\,.
The data points for both system sizes cluster close to the lower of these
two lines in the whole multi-droplet region. However, in the strong-field
part of the region the deviations of the $L$=720 data
from the line are considerably larger than
the one-$\sigma_\Lambda$ error bars, leading to unacceptably small
$\chi^2$-probabilities in a weighted least-squares fit.
These deviations may possibly indicate the presence of small but
statistically significant corrections to the droplet-theory result.
To obtain an acceptable fit for $L$=720
we therefore successively eliminated the data point with the largest value of
$|H|$ until the $\chi^2$-probability $Q$ stabilized at a reasonable
value greater than 0.1. The resulting fit, which includes twelve points
in the field interval 0.03$<$$|H|$$<$0.15, gives
$b \! + \! c \! = \! 1.95(8)$ and $\Xi(0.8T_{\rm c})$=0.282(8) with $Q$=0.41,
consistent with the parameters used to draw the solid straight lines.

Only the smaller system penetrates fully into the single-droplet region in the
field range for which data could be obtained with a reasonable amount of
computer time. As seen from the figure, the agreement
between the $L$=128 MC data and the solid line is also good
in the single-droplet region, even though the error bars are larger there
than in the multi-droplet region.

The dotted straight lines in the figure correspond to Eq.~(\ref{eq13})
with $b \! + \! c \! = \! 3$, the value expected from the discussion following
Eq.~(\ref{eq14}), and the exact $\Xi(0.8T_{\rm c})$.
The agreement with the data is far inferior to the the solid lines.
Since the field-theoretical
prediction of $b \! = \! 1$ has been well confirmed by
several numerical methods
\cite{HARR84,CCAG93,CCAG94,JACU83,PERI84,LW80,WALL82},
we believe the disagreement between the predicted and observed values of
$b$+$c$ must be ascribed to $c$,
the exponent giving the field dependence of the kinetic prefactor.
(An alternative explanation, that the field dependence of the droplet growth
rate, $v_0 \! \sim \! |H|$, might be wrong, is probably ruled out since
this only would affect the fit in the multi-droplet region.) We therefore
conclude that our numerical results agree well with the predictions of
droplet theory, but with $c$$\approx$1 instead of $c$=2.
We believe the unexpected value of $c$
is an expression of the nonuniversality of the kinetic prefactor, and
that it is a consequence of the sequential MC update scheme.
This view is supported by the simulations with
random updates, which we report at the end of this section.

In the strong-field region, droplet theory is not expected to
be applicable. For fields beyond
$H_{\rm MFSP}(0.8T_{\rm c}) \! \approx \! 0.33$, $\Lambda_{\rm eff}$ indeed
exhibits pronounced oscillations with $H$, the first of which
can be seen in the right-hand portion of Fig.~(\ref{fig2}).
Further study at these extreme fields is left for future work.

The crossover between the multi-droplet and single-droplet
regions is clearly seen in Fig.~\ref{fig2}
as a jump in $\Lambda_{\rm eff}$. However, it is
difficult to determine the corresponding dynamic spinodal field
$H_{\rm DSP}$ directly from this figure with sufficient
accuracy to verify Eq.~(\ref{eq15}).
Instead, we consider the relative standard
deviation $r$, which is shown in Fig.~\ref{fig3} on a
logarithmic scale {\it vs.}\ $1/|H|$.
The error bars in the figure are estimated by standard error-propagation
methods as
$$
\sigma_r =
\frac{r}{\sqrt{2(n \! - \! 1)}}
                     \left( 1 + \frac{2(n \! - \! 1)}{n} r^2 \right)^{1/2} \;.
$$
In agreement with our predictions, $r$
crosses over from the behavior described by Eq.~(\ref{eq10b}) for large
$|H|$ to $r$$\approx$1 for smaller $|H|$.
We take as our estimate for
$H_{\rm DSP}$ the field $H_{1/2}$ for which $r$=0.5. This field
is determined for each value of $L$ from the crossing of a weighted
least-squares fit to $\ln r$ in the linear region of Fig.~\ref{fig3} with
the horizontal line $r$=0.5. The resulting
estimates are shown in Fig.~\ref{fig4}a as $1/H_{1/2}$ {\it vs.}\ $L$
on a logarithmic scale, with error bars estimated from
those in Fig.~\ref{fig3} by standard error-propagation methods.
The solid curve in Fig.~\ref{fig4}a,
which agrees well with the data, represents the
crossover condition $L \! \propto \! R_0$, where $R_0$ is given by
Eq.~(\ref{eq7b}) with $b \! + \! c \! = \! 2$
in agreement with the fits in Fig.~\ref{fig2}, and with the proportionality
constant adjusted to approximately minimize the weighted sum of squares.
For reference a dashed line with the asymptotic slope $3/\Xi(0.8T_{\rm c})$,
expected from Eq.~(\ref{eq15}), is also included.
It was obtained by using the same proportionality constant
as in the solid-curve fit and setting the power-law prefactor equal to unity
in Eq.~(\ref{eq7b}).
The values of $H_{1/2}$ for $L$=128 and~720
are also indicated by arrows in Fig.~\ref{fig2}.
The asymptotic $L$ dependence of the lifetime at the DSP, given by
Eq.~(\ref{eq15a}), is illustrated in Fig.~\ref{fig4}b.
For each $L$ this
lifetime was obtained by interpolation between the two closest field
values bracketing $H_{1/2}$ for which simulations had been performed,
and the uncertainty in the resulting estimate was
obtained from those in Figs.~\ref{fig4}a and~\ref{fig1}
by standard error propagation. The numerical results are
consistent with the analytical prediction.

The crossover field between the single-droplet and coexistence regions,
$H_{\rm THSP}$ of Eq.~(\ref{eq17}), is not easily
observable in our calculation because of the smallness of the numerical
factor
${(d \! - \! 1) \widehat{\Sigma}(0.8T_{\rm c})}/
({2 d k_{\rm B} T m_{\rm s}}) \! = \! 0.382 \, 204...$\,.
Nevertheless, for $L$=64 and cutoff at $m$=0.9
($\phi_{\rm s} \! \approx \! 0.03$) the resulting prediction is
$1/H_{\rm THSP} \approx \! 28.5$. As shown in Fig.~\ref{fig5}, which
corresponds to Fig.~\ref{fig1} except that the cutoff is
$m$=0.9, this estimate agrees well with the numerical data.
A similar crossover was also observed in Ref.~\cite{MCCR78}.
Except for the behavior for $L$=64 at
low fields, and larger uncertainties in general,
Fig.~\ref{fig5} is similar to Fig.~\ref{fig1}.

One can ask several questions regarding the effects of different local
dynamics on the metastable lifetimes.
The difference between the Metropolis dynamics used
here and Glauber dynamics should be minor,
at least for relatively weak fields \cite{MKB73}, and we expect our
main conclusions to remain valid for the latter as well. Perhaps more
surprisingly, simulations with a modified Swendsen-Wang algorithm
\cite{RAY90A,RAY90B,SWEN87} also seem to agree qualitatively
with the droplet-theory predictions.

The data presented above were obtained with a
sequential-update algorithm. This is fast and
well-suited to the architecture of the {\it m}-TIS2 computer, but it is
clearly an unrealistic depiction of any physical dynamics. To estimate the
effects of the update scheme
we also simulated the $L$=128 and~720 systems using the same Metropolis
MC algorithm as in the sequential-update study, but with updates at randomly
chosen sites. This study was conducted entirely
on heterogeneous workstation clusters.

The most noticeable difference between the two update schemes is that it takes
a considerably larger number of MCS to reach a given cutoff
magnetization with random updates than with sequential ones. This effect
and its nontrivial field dependence are illustrated in Fig.~\ref{fig7},
which shows the ratio between the random and sequential lifetimes,
$\langle t_{\rm R}(m$=0.7)$\rangle / \langle t_{\rm S}(m$=0.7)$\rangle$,
for $L$=128 and~720.
Since the universal part of the field dependence must cancel in this ratio,
its dominant field dependence is obtained
from Eqs.~(\ref{eq7a}) and~(\ref{eq12}) as
\begin{equation}
\label{eq22}
\frac{\langle t_{\rm R}(\phi_{\rm s}) \rangle}
                                     {\langle t_{\rm S}(\phi_{\rm s}) \rangle}
\propto
\left\{ \begin{array}{ll} |H|^{-\frac{c_{\rm R} - c_{\rm S}}{d+1}}
                                       & \mbox{~in the multi-droplet region} \\
                          |H|^{- ( c_{\rm R} - c_{\rm S})}
                                       & \mbox{~in the single-droplet region,}
            \end{array}\right.
\end{equation}
with proportionality constants that depend only on $T$. Here $c_{\rm R}$
and $c_{\rm S}$ are the nonuniversal
kinetic-prefactor exponents for random and sequential updates, respectively.
This pure power-law behavior is only seen for $L$=720 in the weak-field
portion of the multi-droplet region, indicating that additional
nonuniversal effects distinguish between
the behaviors of the two update schemes for stronger fields.
Following the procedure described in the discussion of Fig.~\ref{fig2}, we
obtained a weighted least-squares fit to the $L$=720 data, including
eight points in the field interval 0.032$\leq$$|H|$$\leq$0.09.
This fit, which is shown as a solid line in the figure,
yields $(c_{\rm R}$$-$$c_{\rm S})$=1.24(3) with $Q$=0.25.
In view of the very short fitting interval, this
estimate should only be taken as an indication that the
difference between the two values of $c$ is probably on the order of unity.

A somewhat puzzling feature in Fig.~\ref{fig7} is the fall-off of the
ratio for $L$=128 in the single-droplet region. We believe this may be due to
the proximity of the THSP for this relatively small system size.

In Fig.~\ref{fig6},
which corresponds to and should be compared with Fig.~\ref{fig2},
we show the effective slope
$\Lambda_{\rm eff}$ from Eq.~(\ref{eq13}) for the random-update case.
The estimated $H_{1/2}$ for both $L$=128 and~720
(indicated by opposing arrows in the figure)
agree to within the statistical error with the estimates for
sequential updates.
In the weak-field portion of the multi-droplet region the effective slopes
follow the lower dotted line, which has the same meaning as in Fig.~\ref{fig2}.
Two major
differences from the sequential-update case are apparent. Firstly, the
field interval in which $\Lambda_{\rm eff}$ seems to follow a linear
approach to the exact $\Xi(0.8T_{\rm c})/3$ is narrower, as one would expect
from the behavior of the lifetime ratios shown in Fig.~\ref{fig7}.
Secondly, the fact that the effective slopes cluster around the {\it dotted}
line indicates that the nonuniversal prefactor exponent $c$ is close to two,
as expected for dynamics that can be described by a
Fokker-Planck equation \cite{LANG68,LANG69,GNW80}.
Following the procedure described above, we obtained
a weighted least-squares fit to the $L$=720 data, including nine points
in the interval 0.032$<$$|H|$$<$0.11. The resulting parameter estimates were:
$\Xi(0.8T_{\rm c})$=0.259(11) and $c$=2.43(15) with $Q$=0.25.
In view of the short fitting interval we consider that this estimate
is consistent with the expected exact values.
For $|H|$$>$0.11 the deviation of $\Lambda_{\rm eff}$ from the dotted straight
line takes the form of a smooth, slow oscillation whose amplitude is much
larger than the statistical errors.

In summary, our results indicate that the main difference in the nonuniversal
part of the field dependence of the metastable lifetimes between the
sequential and random update schemes lies in the kinetic-prefactor exponent
$c$. For sequential updates we find $c$$\approx$1, whereas for random
updates $c$$\approx$2, in agreement with the theoretical expectation $c$=2.
In addition to the difference in $c$
for weak fields, the quantitative differences between
the lifetimes in the two update schemes become progressively larger with
increasing $|H|$.

\section{Discussion}
\label{sec5}

We have demonstrated
that the metastable lifetimes for impurity-free kinetic Ising
models with nearest-neighbor interactions and local dynamics, in
weak to moderate unfavorable applied fields, exhibit
system-size and field dependences that can be explained by
a field-theoretical droplet model of homogeneous nucleation
\cite{LANG67,LANG68,LANG69,GNW80}.
The most significant determining factor for the lifetime
is the free energy of a critical droplet of the
stable phase, which in a wide field interval
leads to an exponential dependence on the inverse
applied field. This free energy is independent of the
details of the local dynamics, and it can
be accurately calculated by combining a Wulff construction with the exact
zero-field equilibrium surface tension.

Our numerical estimates for the lifetimes, based on
Monte Carlo simulations in two dimensions, are in excellent agreement
with the droplet model. This agreement includes
both the aforementioned exponential
dependence on the inverse field, and multiplicative power-law prefactors.
There are two independent contributions to these prefactors:
a universal exponent $b$, equal to
unity for $d$=2, and a nonuniversal exponent $c$, which depends on the
details of the dynamics.
Our results indicate that with MC updates at randomly chosen sites,
$c$ is close to two, which is the exact result
for dynamics described by a Fokker-Planck equation \cite{LANG68,LANG69}.
With sequential updates we found
$c$$\approx$1. It is tempting to conjecture that the exact value in
this case is unity, but at present we have no theoretical arguments to
substantiate this suggestion.
For stronger fields we found additional, dynamics-dependent
deviations from the droplet-theory predictions for $\Lambda_{\rm eff}$.

We have identified four different field regions, in which the
decay proceeds through different excitations.
In order of increasingly strong
unfavorable field $|H|$, these were the ``coexistence region'', characterized
by subcritical fluctuations on the scale of the system volume; the
``single-droplet region'', characterized by decay via a single critical
droplet; the ``multi-droplet region'', characterized by decay via a finite
density of droplets; and the ``strong-field
region'', in which the droplet picture is inappropriate.
The crossover fields between these regions,
\begin{equation}
\label{eq20}
[ H_{\rm THSP} \! \sim \! L^{-1} ]
< [ H_{\rm DSP} \! \sim \! (\ln L)^{-\frac{1}{d-1}} ]
< [ H_{\rm MFSP} \sim L^0 ] \, ,
\end{equation}
are accurately predicted by droplet theory.
The different regions and crossover fields are illustrated in
Fig.~\ref{fig0}.
We believe the slow, logarithmic vanishing with system size
of the dynamic spinodal field $H_{\rm DSP}$, which separates the
single-droplet and multi-droplet regions, is of particular significance. This
convergence is so slow, especially in three and higher
dimensions, that systems that
are rightfully considered macroscopic in terms of their equilibrium properties
may nevertheless possess a measurable single-droplet region in which the
metastable state is exceedingly long-lived on the average.

\acknowledgments

We gratefully acknowledge discussions with J.~Lee, M.~A.\ Novotny,
B.~M.\ Gorman, and C.~C.~A.\ G{\"u}nther, and helpful
comments on an earlier version of this paper by D.~Stauffer.
The numerical values for $\widehat{\Sigma}(0.8T_{\rm c})$ and
$H_{\rm MFSP}(0.8T_{\rm c})$ were provided by C.~C.~A.\ G{\"u}nther.
P.~A.~R.\ greatly enjoyed the kind hospitality at Kyoto University.
This study was
supported by a Grant-In-Aid for Scientific Research on Priority Area
``Computational Physics as a New Frontier in Condensed Matter Research''
from the Ministry of Education, Science, and Culture of Japan,
and by National Science Foundation Grant No.~DMR-9013107.
It was also supported by Florida State University through the Supercomputer
Computations Research Institute (Department of Energy Contract No.\
DE-FC05-85ER25000) and the Center for Materials Research and Technology.


\begin{figure}
\caption[]{
Schematic sketch showing the field dependence of
the average metastable lifetime for a $d$-dimensional kinetic Ising model
on a logarithmic scale.
The dynamic spinodal point is indicated by the solid vertical line labelled
DSP. To its left is the deterministic region, and to its right is the
stochastic region.
The deterministic region is divided into the strong-field (SF) and
multi-droplet (MD) subregions by the mean-field spinodal, indicated by the
dashed vertical line labelled MFSP.
The stochastic region is divided into the single-droplet (SD) and
coexistence (CE) subregions by the thermodynamic spinodal, indicated by the
dashed vertical line labelled THSP.
This figure is based on the analytical and
numerical results presented in this paper, but it is not drawn to scale for
a specific system size. In particular, the extent of the single-droplet
region is greatly reduced, relative to the multi-droplet region.
}
 \label{fig0}
 \end{figure}

\begin{figure}
\caption[]{
Mean metastable lifetime for a two-dimensional kinetic Ising model
at $T$=0.8$T_{\rm c}$, estimated as
average first-passage times to $m$=0.7 ($\phi_{\rm s} \! \approx \! 0.13$).
The simulation was performed by the Metropolis MC algorithm with sequential
updates.
The lifetimes are given in
units of Monte Carlo steps per spin (MCS) and are
shown on a logarithmic scale {\it vs.}\ $1/|H|$.
The symbols indicate data for system sizes $L$=64 ($+$), 128 ($\bigcirc$),
256 ($\times$), 400 ($\Box$), and 720 ($\Diamond$),
all obtained with a special-purpose {\it m}-TIS2 computer, and
$L$=128 ($\bigotimes$) and~720 (filled diamonds),
obtained on heterogeneous workstation clusters.
The solid curves are merely guides to the eye.
The vertical arrow indicates the inverse field at which droplet theory is
expected to break down, $1/H_{\rm MFSP} \! \approx \! 3.0$.
The $L$-dependent data in the upper right-hand
sector lie in the single-droplet region,
the $L$-independent data in the central portion of the figure lie in the
multi-droplet region, and
the $L$-independent data in the lower left-hand
sector belong to the strong-field region.
The dashed line above the data has a slope of 0.36,
estimated from the four points for $L$=64 with $1/|H| \! \ge \! 20$,
and the dashed line below the data has 1/3 this slope.
The standard deviations in the average lifetimes
range from on the order of the
symbol size for the weakest fields shown, to two orders of magnitude smaller
for the strongest fields, and error bars are therefore not shown.
See details in Sec.~\ref{sec4}.
}
 \label{fig1}
 \end{figure}

 \begin{figure}
 \caption[]{
Finite-difference estimates for the effective slope
$\Lambda_{\rm eff}$ from Eq.~(\ref{eq13}), shown {\it vs.}\ $|H|$.
The parameters and symbols are the same as in Fig.~\ref{fig1}.
For clarity only data for $L$=128 ($\bigcirc$ and $\bigotimes$)
and~720 (empty and filled diamonds) are
included, and error bars are shown only where larger than the symbol size.
The solid straight lines correspond to
Eq.~(\ref{eq13}) with $b \! + \! c \! = \! 2$ and intersect the vertical axis
at the exact $\Xi(0.8T_{\rm c})$ and $\Xi(0.8T_{\rm c})/3$, respectively.
The lower solid line gives a good fit to the data in the multi-droplet
region for both system sizes.
Only $L$=128 penetrates fully into the single-droplet region, where
the agreement with the theoretical prediction (upper solid line) is also good.
The dotted lines, which correspond to
$b \! + \! c \! = \! 3$, do not fit the data as well.
The crossover between the multi- and single-droplet
regions is seen as a jump in $\Lambda_{\rm eff}$.
The vertical arrows with
horizontal error-bar ``feathers'' indicate the positions of the
estimator $H_{1/2}$ for the crossover field $H_{\rm DSP}$,
obtained from Fig.~\ref{fig4}a. The left pair of arrows corresponds to
$L$=720, and the right pair to $L$=128.
The vertical arrow marked $H_{\rm MFSP}$ marks the $L$-independent
field at which droplet theory is expected to break down.
See details in Sec.~\ref{sec4}.
}
 \label{fig2}
 \end{figure}

 \begin{figure}
 \caption[]{
The relative standard deviation $r$ for $t(m$=0.7),
shown on a logarithmic scale {\it vs.}\ $1/|H|$.
The parameters and symbols are the same as in Fig.~\ref{fig1}.
The behavior of $r$ crosses over from the approximate
straight line described by
Eq.~(\ref{eq10b}) in the multi-droplet region to $r \! \approx \! 1$
in the single-droplet region.
The inclined solid lines are weighted least-squares fits
to the data in the region $1/|H|$$\ge$10, $r$$\le$0.6. Their average
effective slope is 0.13(1),
which is intermediate between $\Xi(0.8T_{\rm c})$ and
$\Xi(0.8T_{\rm c})/3$.
The estimates $H_{1/2}$ for $H_{\rm DSP}$ are found
where these lines cross the horizontal dashed line at $r$=0.5.
These estimates
are also shown {\it vs.}\ $L$ in Fig.~\ref{fig4}a and are indicated by arrows
in Fig.~\ref{fig2}.
See details in Sec.~\ref{sec4}.
}
 \label{fig3}
 \end{figure}

 \begin{figure}
 \caption[]{
Inverse field and mean lifetime at the DSP.
(a): The estimates $1/H_{1/2}$ for $1/H_{\rm DSP}$ as obtained from
Fig.~\ref{fig3}, shown  {\it vs.}\ $L$ on a logarithmic scale.
The solid curve corresponds to $L$$\propto$$R_0$
with $b \! + \! c \! = \! 2$ in Eq.~(\ref{eq7b}) and with the
proportionality constant adjusted to minimize the weighted sum of squares.
The dashed line indicates the asymptotic slope $3/\Xi(0.8T_{\rm c})$.
(b): The average lifetimes at $1/H_{1/2}$. The solid straight
line is a weighted least-squares fit. The figure supports the
asymptotic behavior given in Eq.~(\ref{eq15a}),
$\langle t \; {\rm at} \; H_{\rm DSP} \rangle \propto L \ln L$.
See details in Sec.~\ref{sec4}.
}
 \label{fig4}
 \end{figure}

 \begin{figure}
 \caption[]{
Mean metastable lifetimes at $T$=0.8$T_{\rm c}$, estimated as
average first-passage times to $m$=0.9 ($\phi_{\rm s} \! \approx \! 0.03$).
The parameters and symbols are otherwise the same as in Fig.~\ref{fig1}.
Except for somewhat larger uncertainties and the behavior for $L$=64 at
low fields, this figure is similar to Fig.~\ref{fig1}.
The guide-to-the-eye curves are dotted, except for
$L$=64, which has been drawn solid to emphasize the
behavior for this smallest system. It crosses over between the single-cluster
and coexistence regions at a $\phi_{\rm s}$-dependent
$H_{\rm THSP} \! \approx \! 28.5$ (marked by the opposing arrows),
as discussed in Sec.~\ref{sec4}.
}
 \label{fig5}
 \end{figure}

 \begin{figure}
 \caption[]{
The ratio $\langle t_R \rangle / \langle t_S \rangle$ between the mean
lifetimes obtained with random and sequential MC updates for $L$=128
and~720.
The symbols and parameters are the same as in Figs.~\ref{fig2} and~\ref{fig6}.
The two pairs of opposing arrows with
horizontal error-bar ``feathers'' mark $H_{1/2}$ for the two system sizes.
The solid straight line in this log-log plot
is a weighted least-squares fit to Eq.~(\ref{eq22})
in the weak-field part of the multi-droplet region for $L$=720.
See details in Sec.~\ref{sec4}.
}
 \label{fig7}
 \end{figure}

 \begin{figure}
 \caption[]{
Finite-difference estimates for the effective slope
$\Lambda_{\rm eff}$ from Eq.~(\ref{eq13}) for random MC updates with $L$=128
and~720,
shown {\it vs.}\ $|H|$. The figure is analogous to Fig.~\ref{fig2}, and the
symbols and parameters are the same as in that figure.
The behavior is similar to the sequential-update case, except that the
prefactor exponent $b \! + \! c \! \approx \! 3$, and the field interval
in which $\Lambda_{\rm eff}$ seems to follow a linear approach to the
exact $\Xi$/3 is narrower than in Fig.~\ref{fig2}.
See details in Sec.~\ref{sec4}.
}
 \label{fig6}
 \end{figure}


\begin{references}
\bibitem{ZIA82X}
Our quantity $\widehat{\Sigma}(T)$ is identical to $d W^{1/d}$ in Eq.~(3)
of Ref.~\cite{ZIA82}.
\end{references}
\end{document}